\documentstyle[aps,preprint,psfig,epsf]{revtex}
\tighten
\begin{document}
\draft
%\twocolumn[
\hsize\textwidth\columnwidth\hsize\csname @twocolumnfalse\endcsname

\title{Towards Bose-Einstein Condensation of  
Electron Pairs: Role of Schwinger Bosons}
\author{Gang Su$^{\ast}$ and Masuo Suzuki$^{\dag}$}
\address{ Department of Applied Physics, Faculty of Science,
 Science University of Tokyo\\
1-3, Kagurazaka, Shinjuku-ku, Tokyo 162, Japan}
%\date{\today}

\maketitle

\begin{abstract}
 
It can be shown that the bosonic degree of freedom 
of the tightly bound on-site electron
pairs could be separated as Schwinger bosons.  This is
implemented by projecting the whole Hilbert space into the
Hilbert subspace spanned by states of two kinds 
of Schwinger bosons (to be called {\em binon} and 
{\em vacanon}) subject to a constraint 
that these two kinds of bosonic quasiparticles 
cannot occupy the same site. 
We argue that a binon is actually a kind of quantum fluctuations of 
electron pairs, and a vacanon corresponds to
a vacant state. These two bosonic quasiparticles may 
be responsible for the Bose-Einstein 
condensation (BEC) of the system associated with electron pairs. 
These concepts are also applied to the attractive Hubbard model 
with strong coupling, showing that it is quite useful. 
The relevance of the present arguments to the existing theories
associated with the BEC of electron pairs is briefly commented. 
 
\end{abstract}

%]

\section*{Introduction}

The discovery of pseudogap phenomenon in the normal state of
underdoped cuprate oxides renews
the interest of exploiting the Bose-Einstein condensation (BEC) 
of electron pairs. One of theories attributes this unusual phenomenon
to the BEC of the pre-formed singlet pairs under such an assumption that
a smooth crossover from the BEC to the BCS regimes can emerge with
increasing carrier concentration\cite{ue1}. 
There are also other theories related to the
pre-formed electron pairs (see, e.g. Ref.\cite{ran} for a review)
or bipolarons\cite{jan} in cuprates. On the other hand,
the evolution from the BCS theory to the BEC regime has been discussed long  
time ago by Eagles\cite{eag}, Leggett\cite{leg}, Nozi\`eres and Schmitt-Rink
\cite{noz}, and others (see, e.g. Ref.\cite{ran1} for a review).
A key ingredient is to modify the electron-phonon interacting potential
by plugging a two-particle scattering length in the BCS reduced
Hamitonian so that  a crossover parameter $\xi$  
proportional to the inverse of the scattering length, is introduced,
which results in
the crossover from the BCS condensation ($\xi \to -\infty$) 
of the Cooper pairs to
the BEC ($\xi \to +\infty$) of tightly bound 
electron pairs. The chemical potential is zero at the transition 
point. Note that the Cooper pairs are formed in 
the weak coupling limit in momentum space, while
tightly bound electron pairs are formed in the strongly attractive 
coupling limit in real space. 

With regard to the aforementioned theories, 
people intuitively believe that in the strong
coupling limit the tightly bound electron pairs can be
taken as bosons, each of which may be constituted by two 
fermions (e.g. electrons) through certain mechanism, and they
can undergo the BEC below a certain temperature. 
However, whether or not the tightly bound electron pairs can really  
undergo the BEC {\em directly}, is an old but fundamental 
(and nontrivial) problem, 
particularly as we know that the bound electron pairs are not exactly 
bosons\cite{note00}. If not, what are the corresponding bosons 
in these theories? 
On the other hand, in the boson-fermion model for
superconductivity\cite{jan} a localized boson (or a bipolaron) 
which was also regarded as a kind of tightly bound electron pair, 
was simply replaced by a hard-core boson. What is the underlying 
physics behind such a simple replacement?
In this paper, we shall try to address these questions. 
Our result appears to show that the fluctuations of the
tightly bound electron pairs play a central role. In the following,
we shall first show why the tightly bound electron pairs could not
undergo the BEC {\em directly} in momentum space using the standard
argument, and then introduce two kinds of Schwinger bosons ({\em binon} 
and {\em vacanon}) and argue that a binon is a kind of fluctuations 
of the tightly bound electron pairs and  a vacanon corresponds to a 
vacant state. These two bosonic quasiparticles may be responsible 
for the BEC of the system. These concepts are also applied to the 
attractive Hubbard model in the strong coupling limit, and it appears
that this special system can be exploited on the basis of 
binons and vacanons. Finally, we shall discuss briefly 
the relevance of our argument to the
existing theories concerning the BEC of electron pairs.

\section*{Useful Notions}

The operators  
of a tightly bound electron pair are defined in real space as usual 
in the following\cite{note0}:
\begin{eqnarray}
b_{i} = c_{i\downarrow}c_{i\uparrow}, ~~~b_{i}^{\dag}=(b_{i})^{\dag},
\label{boson}
\end{eqnarray}
where $c_{i\uparrow}$$(c_{i\downarrow})$ is 
the annihilation operator of an electron with
up(down)-spin at site $i$ obeying the anticommutation relations. 
It is well-known that they satisfy the standard
SU(2) algebra:
%\begin{eqnarray}
$[b_{i}, b_{j}^{\dag}] = 2 b^{0}_{i}\delta_{ij},$ 
$[b_{i}^{\pm}, b_{j}^{0}] = \pm b_{i}^{\pm}\delta_{ij},$ 
%\label{commu-i}
%\end{eqnarray}
with $b^{0}_{i} = \frac12 (1 - n_{i})$ and 
$n_{i} = c_{i\uparrow}^{\dag}c_{i\uparrow} + 
c_{i\downarrow}^{\dag}c_{i\downarrow}$ the number operator of 
electrons at site $i$, as well as the bosonic
property $[b_{i}, b_{j}] = [b_{i}^{\dag}, b_{j}^{\dag}] =0$.  
Owing to Pauli's exclusion principle, the
tightly bound electron pairs exhibit the hard-core property, 
satisfying
\begin{eqnarray}
b_{i}^2 = {b_{i}^{\dag}}^2 =0.
\label{hc-i}
\end{eqnarray}
Moreover, it can be observed that there exists an important property 
(i.e., an anticommutator):
\begin{eqnarray}
\{b_{i}, b_{i}^{\dag}\} = 4 (b^{0}_{i})^2,
\label{commu-fm}
\end{eqnarray}
implying that a tightly bound electron pair has fermionic characters.
As the BEC takes place only in momentum space, we need to consider their
Fourier transforms. The hermitian property ($b_{i}^{\dag}=(b_{i})^{\dag}$)
and the consistency require that the Fourier 
transforms of the operators for the tightly bound electron pairs can 
only be defined through the following forms:
\begin{eqnarray}
& & b_{i} = \frac{1}{M}\sum_{{\bf k}} e^{i {\bf k}\cdot {\bf r}_{i}} 
{\tilde b}_{\bf k}, ~~~
{\tilde b}_{\bf k} = \sum_{i}e^{-i {\bf k}\cdot {\bf r}_{i}}b_{i}  
\nonumber \\
& & b_{i}^{\dagger} = (b_{i})^{\dag} = \frac{1}{M}\sum_{{\bf k}} 
e^{-i {\bf k}\cdot {\bf r}_{i}} {\tilde b}_{\bf k}^{\dag}, ~~~
{\tilde b}_{\bf k}^{\dag} = \sum_{i}e^{i {\bf k}\cdot 
{\bf r}_{i}}b_{i}^{\dag}
  \nonumber \\
& & b_{i}^{0} = \frac{1}{M}\sum_{{\bf k}} e^{i {\bf k}\cdot {\bf r}_{i}} 
{\tilde b}_{\bf k}^{0}, ~~~
{\tilde b}_{\bf k}^{0} = \sum_{i}e^{-i {\bf k}\cdot {\bf r}_{i}}b_{i}^{0},
\label{fourier-b}
\end{eqnarray}
where $M$ is the number of lattice sites, and  Kronecker's 
delta function is 
defined by $\delta_{ij}
= \frac{1}{M}\sum_{{\bf k}} e^{i {\bf k}\cdot ({\bf r}_{i}-{\bf r}_{j})}$.
It should be noted that the operator 
${\tilde b}_{\bf k}$ differs from the Cooper 
pair operator $c_{-{\bf k}\downarrow}c_{{\bf k}\uparrow}$ (see below).
The commutation relations of the Fourier transforms are of the form:
%\begin{eqnarray}
$ [{\tilde b}_{\bf k}, {\tilde b}_{\bf k'}^{\dag}] = 
2 {\tilde b}_{{\bf k}-{\bf k}'}^{0},$
$[{\tilde b}_{\bf k}, {\tilde b}_{\bf k'}^{0}] = 
- {\tilde b}_{{\bf k}+{\bf k}'},$
$[{\tilde b}_{\bf k}^{\dag}, {\tilde b}_{\bf k'}^{0}] =  
{\tilde b}_{{\bf k}-{\bf k}'}^{\dag}, $
$ [{\tilde b}_{\bf k}, {\tilde b}_{\bf k'}] =
[{\tilde b}_{\bf k}^{\dag}, {\tilde b}_{\bf k'}^{\dag}] = 0.$
%\label{fourier-commu}
%\end{eqnarray} 
Considering the Fourier transform of the electron operator,
$c_{i\sigma} =(1/\sqrt{M}) \sum_{{\bf k}} \exp (i {\bf k}\cdot 
{\bf r_{i}}) c_{{\bf k}\sigma}$ ($\sigma =\uparrow, \downarrow$), one
may find the relationship between operators $\{{\tilde b}_{\bf k}\}$ 
and electron operators:
%\begin{eqnarray}
${\tilde b}_{\bf k} = \sum_{\bf q} c_{{\bf q}+\frac{{\bf 
k}}{2}\downarrow} c_{-{\bf q}+\frac{{\bf k}}{2}\uparrow}$, 
${\tilde b}_{\bf k}^{\dag}  = ({\tilde b}_{\bf k})^{\dag}$, and 
${\tilde b}_{\bf k}^{0} = \frac12 \sum_{\bf q} (\delta_{{\bf k},0}
- c_{{\bf q}-\frac{{\bf k}}{2}\downarrow}^{\dag}
c_{{\bf q}+\frac{{\bf k}}{2}\downarrow} 
- c_{{\bf q}-\frac{{\bf k}}{2}\uparrow}^{\dag}
c_{{\bf q}+\frac{{\bf k}}{2}\uparrow})$.
%\label{fourier-elec}
%\end{eqnarray}
Clearly, the tightly bound electron pair in momentum space consists of
the superposition of two electrons with different momenta and is 
spin-singlet. Unlike the standard Cooper pair, the total momenta of 
the two pairing electrons here is nonzero.
It is interesting to mention that the hard-core property
like Eq. (\ref{hc-i}) no longer holds in momentum space, namely, 
%\begin{eqnarray}
${\tilde b}_{\bf k}^2 = \sum_{{\bf q},{\bf q}'({\bf q}\neq {\bf q}')}
 c_{{\bf q}+\frac{{\bf 
k}}{2}\downarrow} c_{-{\bf q}+\frac{{\bf k}}{2}\uparrow} 
c_{{\bf q}'+\frac{{\bf 
k}}{2}\downarrow} c_{-{\bf q}'+\frac{{\bf k}}{2}\uparrow} \neq 0$,
%\end{eqnarray}
and the anticommutator like Eq. (\ref{commu-fm}) will not remain true
in momentum space,
suggesting that the fermionic properties which exist in real space do
not appear in momentum space. Therefore, in 
contrast to the Cooper pair $c_{-{\bf k}\downarrow}c_{{\bf k}\uparrow}$
which possesses exactly the same commutation relations as 
the tightly bound electron pair in
real space, the latter
may purely show bosonic behavior in momentum space,  which
appears in turn to suggest that the BEC of the tightly bound electron
pairs in momentum space be possible.  

\section*{Tightly Bound Electron Pairs in Momentum Space}

Now,
let us examine if the BEC of the noninteracting tightly bound 
 electron pairs in momentum space can really occur. By noting
 the fact $\sum_{{\bf k}}{\tilde b}_{\bf k}^{\dag}{\tilde b}_{\bf k}/M
 = \sum_{i}b_{i}^{\dag}b_{i}$, we define the number operator
 per volume, ${\hat {\cal N}_{\bf k}}$,
 of the tightly bound electron pairs in momentum space as
${\hat {\cal N}_{\bf k}} = {\tilde b}_{\bf k}^{\dag}{\tilde b}_{\bf k}/M$.
The eigenvalues of ${\hat {\cal N}_{\bf k}}$ can
be argued to be non-negative integers in the infinite volume of the system. 
This may be reached by observing the following facts. 
First, the eigenvalues of ${\hat {\cal N}_{\bf k}}$
are non-negative, because for any eigenstate we can write $\langle m| 
{\tilde b}_{\bf k}^{\dag}{\tilde b}_{\bf k}|m \rangle = \sum_{n}
\langle m|{\tilde b}_{\bf k}^{\dag}|n\rangle|^2 \geq 0$ where 
$\{|n\rangle\}$ is a complete orthonormal set of eigenstates. Second, 
we can construct the orthonormal set of eigenstates of
${\hat {\cal N}_{\bf k}}$ as
$|n,{\bf k}\rangle = [(M-n)!/(M!n!)]^{1/2} ({\tilde b}_{\bf k}^{\dag})^n
|0\rangle$ satisfying $\langle n', {\bf k}'|n,{\bf k}\rangle =
\delta_{n,n'} \delta_{{\bf k},{\bf k}'}$, where the Fock vacuum 
$|0\rangle$ satisfies $c_{{\bf k}\sigma}|0\rangle =0$. It is 
interesting to point out that the state $|n,{\bf k}\rangle$ possesses
the on-site off-diagonal long-range order (ODLRO) in the two-particle
reduced density matrix when $n$ is on the order of $M$. One may easily 
examine that $ {\hat {\cal N}_{\bf k}} |n,{\bf k}\rangle =
n(M-n+1)/M|n,{\bf k}\rangle$. Since $n$ ($\leq M$) 
is integers, it is possible that, in the infinite volume (i.e., $M \to 
\infty$), the eigenvalues ${\cal N}_{\bf k}$ in the given ${\bf k}$ 
state can be taken as ${\cal N}_{\bf k} = 0, 1, 2, 3, \cdots, \infty$
with the degeneracy being $2$.
For the noninteracting system of the tightly bound electron pairs
with the single-particle dispersion
$\epsilon_{{\bf k}}$, it seems that one can get the average
occupation number per volume $\langle {\hat {\cal N}_{\bf k}} 
\rangle$ in the following way:
\begin{eqnarray}
\langle {\hat {\cal N}_{\bf k}} \rangle = \frac{2\sum_{{\cal N}_{\bf 
k}=0}^{\infty} e^{-\beta (\epsilon_{{\bf k}}-\mu_b) {\cal N}_{\bf k}}
{\cal N}_{\bf k}}{\sum_{{\cal N}_{\bf 
k}=0}^{\infty} e^{-\beta (\epsilon_{{\bf k}}-\mu_b) {\cal N}_{\bf k}}}
= \frac{2}{e^{\beta (\epsilon_{{\bf k}}-\mu_b)}-1},
\label{average}
\end{eqnarray}
with $\mu_b$ the chemical potential of the tightly bound electron pairs.
If this formula is correct, the BEC will appear in this system, like the 
usual ideal Bose gas. Unfortunately, when one looks carefully at 
Eq. (\ref{average}), one may find it incorrect. The reasoning is as follows.
Equation (\ref{average}) implies that the Hamiltonian for this noninteracting
system is of the form  $(1/M)\sum_{{\bf k}}(\epsilon_{{\bf k}}-\mu_b)
{\tilde b}_{\bf k}^{\dag}{\tilde b}_{\bf k}$ which looks apparently like
the conventional noninteracting Bose system. However, when considering
the fact that the operator ${\hat {\cal N}_{\bf k}}$ does not commute with 
the Hamiltonian of this form owing to $[{\hat {\cal N}_{\bf k}},
{\hat {\cal N}_{{\bf k}'}}] \neq 0$ for ${\bf k} \neq {\bf k}'$, 
suggesting that the Hamiltonian 
cannot be diagonalized by the eigenstates of ${\hat {\cal N}_{\bf k}}$, one
may observe that the first equality of Eq. (\ref{average}) is not
justified. This could also be easily understood if we loosely treat 
the bound electron pairs as pseudospins and thereby the system
as the XY model.
Therefore, it might be reasonable to believe that 
the noninteracting tightly bound electron pairs in momentum 
space could not {\em directly} undergo the BEC
in the usual (or strict) sense that the BEC is a phenomenon for 
the noninteracting Bose systems\cite{note00}.
On the other hand, as advocated by a number of people,
there must be a crossover from the BCS theory to the BEC of   
the tightly bound electron pairs  based on qualitative arguments
associated with approximate calculations.
Naively speaking, this is certainly sound, because if the attractive
interactions between electrons are strong enough, the tightly bound
electron pairs are formed, and the low-lying states of the system are
thus occupied by electron pairs, which might have a macroscopic
occupation at a certain quantum state due to the bosonic
nature of these pairs.
How to reconcile these facts? 
Actually, from our point of view 
the tightly bound electron pairs or Cooper pairs cannot undergo the
BEC themselves directly, but their fluctuations probably can. These
fluctuations can be constructed as genuine bosons through some kind
of nonlinear transformations\cite{note2}, in which Pauli's
exclusion principle seemingly no longer plays a role in the formalism. 
Consequently, owing to the genuinely bosonic statistical 
property of these quasiparticles the BEC of
the fluctuations appear to be possible. 
Emery and Kivelson have recently
presented a nice example, similar in spirit to this argument, 
in which they assumed that the superconducting order parameter
of a metal, with an amplitude and a phase, is complex, and
emphasized the importance of
phase fluctuations in superconductors with small superfluid 
density\cite{ek}. Here we shall offer an alternative   
realisation.

\section*{Binons and Vacanons}
 
We introduce two kinds of bosons by which a tightly bound 
electron pair can be
constituted. To this aim, we define
\begin{eqnarray}
b_{i} = f^{\dag}_{i}a_{i}, ~~~ b_{i}^{\dag} = a_{i}^{\dag}f_{i}, ~~~
b_{i}^{0} = \frac12 (n_{i}^f - n_{i}^a)
\label{trans}
\end{eqnarray}
where the operators $\{f\}$ and $\{a\}$ are of bosons, satisfying 
the standard
commutation relations: $[f_{i}, f_{j}^{\dag}] = [a_{i}, a_{j}^{\dag}]
= \delta_{ij}$ and
$[f_{i},f_{j}]=[f_{i}^{\dag}, f_{j}^{\dag}]=$ 
$[a_{i},a_{j}]=[a_{i}^{\dag}, a_{j}^{\dag}]=0$. 
In addition, they
commute each other, namely $[\{f\}, \{a\}]=0$. 
These bosons are hard-core bosons, obeying
\begin{eqnarray}
(n_{i}^{a})^2 = n_{i}^{a}, ~~~ (n_{i}^{f})^2 = n_{i}^{f},  
\label{fa-hc}
\end{eqnarray}
with $n_{i}^{f} (n_{i}^{a}) = 
f_{i}^{\dag}f_{i} (a_{i}^{\dag}a_{i})$ the number operator 
of the respective bosons.
Furthermore, we impose a constraint on the two bosons:
\begin{eqnarray}
n_{i}^a  n_{i}^f = 0.
\label{nfna=0}
\end{eqnarray}
This condition, 
arising from the consistency
requirement with  Eq.(\ref{commu-fm}), implies that the two bosons cannot
occupy the same site.
The underlying physics behind Eqs. (\ref{fa-hc}) and
(\ref{nfna=0}) is still Pauli's exclusion principle. 
One may check that with the above definitions  
all the conditions imposed on the tightly bound electron pair operators  
are indeed satisfied.
We call the boson "a" as {\em binon} and the boson "f" as {\em vacanon},
which are actually two kinds of Schwinger bosons, because the 
transformation (\ref{trans}) is seemingly similar to the form 
in the context of angular momentum discussed long time ago by Schwinger.
It turns out that the number of the tightly bound electron
pairs maps onto the following
\begin{eqnarray}
b_{i}^{\dag}b_{i} \Rightarrow a_{i}^{\dag}a_{i}.
\label{b->a}
\end{eqnarray}
As a consequence, the total number of electrons, $N_{e} =\sum_{i}n_i$,
becomes $N_{e} = M + N_a - N_f$, where $N_{a(f)} = \sum_{i}n_{i}^{a(f)}$,
is the total number of binons (vacanons).
This reveals that the role of the tightly bound electron
pairs played in the original electron
system is now replaced by that of the new hard-core quasiparticles
(i.e. binons and vacanons) in the new system. The fermionic
property of the electron pairs is reflected by the conditions
(\ref{fa-hc}) and (\ref{nfna=0}). 
As can be seen, this transformation is actually a {\em projection}
procedure, which maps the whole Hilbert space ${\cal V}$ spanned by all 
the states of the tightly bound electron 
pairs onto the Hilbert subspace ${\cal V}_0$ spanned by those
of the new bosons with the constraint that the two kinds of Schwinger
bosons cannot occupy the same site. 
The projection operator ${\cal P}$ can be defined such
that ${\cal P}:$ ${\cal V} \Rightarrow {\cal V}_0$, and 
${\cal P}:$ $b_{i}^{\dag}b_{i} \Rightarrow a_{i}^{\dag}a_{i}$ with
the Hamiltonian
$H \Rightarrow {\tilde H}$. The transformation defined in 
Eq. (\ref{trans}) just gives a possible manifestation of ${\cal P}$.
It is in some sense very similar to the projection made in the context
of the t-J model where the doubly-occupied sites are projected out.
We could remark here that the Bose quasiparticles (binons and vacanons) 
in the system of the tightly bound electron pairs are somewhat in   
analogy with the magnons in magnetic systems. 

The vacanon is an auxiliary Bose field by which the commutators
for operators $\{b_i\}$ are guranteed. 
For the Fock vacuum $|0\rangle$, we have $b_{i}|0\rangle =0$
and $b_{i}^{\dag}|0\rangle \neq 0$. By definition, we observe
\begin{eqnarray}
a_{i}|0\rangle =0, ~~~~f_{i}|0\rangle \neq 0.
\label{af-0}
\end{eqnarray}
This is the intrinsic difference between the two Schwinger 
bosons. The vacanon can
be understood as a kind of empty or vacancy
on the lattice in the sense that it cannot be destructed in
vacuum\cite{note5}.   
Since a tightly bound electron pair carries spin zero 
and electric charge 2,
namely, $s_{i}^z b_{i}^{\dag} =0$ and $n_{i}b_{i}^{\dag}=2 b_{i}^{\dag}$,
we have $s_{i}^z a_{i}^{\dag} =0$ and $n_{i}a_{i}^{\dag}=2 a_{i}^{\dag}$,
where $s_{i}^z$ is the z-component of the spin-1/2 operator, and use has
been made of conditions (\ref{af-0}). This shows that {\em a binon 
also carries spin zero and electric charge 2}. 
It is not difficult to verify that
a vacanon carries both charge and spin zero. 
In this way, the hopping of a tightly
bound electron pair on a lattice from the site $j$ to the site $i$ is 
equivalent to the hopping of a binon from the site $j$ to the site $i$
while accompanying the hopping of a vacanon from the site $i$ to the 
site $j$, as depicted in Fig.1. This makes the separation of
the genuinely charged hard-core bosonic degrees of freedom and
the vacant states originally implied in the system of the
tightly bound electron pairs. Note that the hopping occurs  
only between the occupied and unoccupied sites.
Let us discuss briefly the physical meaning
of the binon. It is readily checked that the following 
commutators hold:
$ [b_{i}, a_{i}] = [b_{i}, f_{i}^{\dag}]= [b_{i}^{\dag}, f_{i}]
=[b_{i}^{\dag}, a_{i}^{\dag}] =0,$  $ [b_{i}, f_{i}] = - a_{i}$, 
$[b_{i}, a_{i}^{\dag}]= f_{i}^{\dag}$,  
$[b_{i}^{\dag}, a_{i}] = - f_{i}$, $[b_{i}^{\dag}, f_{i}^{\dag}]
= a_{i}^{\dag}$.
In momentum space, we have
\begin{eqnarray}
& & a_{{\bf k}}^{\dag} = \frac{1}{M} \sum_{{\bf q}}
({\tilde b}_{{\bf q} +\frac{{\bf k}}{2}}^{\dag} 
f_{-{\bf q} +\frac{{\bf k}}{2}}^{\dag} - 
f_{-{\bf q} +\frac{{\bf k}}{2}}^{\dag}
{\tilde b}_{{\bf q} +\frac{{\bf k}}{2}}^{\dag}),  
\label{a_k} \\
& & f_{{\bf k}}^{\dag} = \frac{1}{M} \sum_{{\bf q}}
({\tilde b}_{{\bf q} -\frac{{\bf k}}{2}} 
a_{{\bf q} +\frac{{\bf k}}{2}}^{\dag} - 
a_{{\bf q} +\frac{{\bf k}}{2}}^{\dag}
{\tilde b}_{{\bf q} -\frac{{\bf k}}{2}}),
\label{f_k}  
\end{eqnarray}
where we have supposed that the Fourier transforms 
of the Schwinger bosons have the same form as electrons.
Since the boson operator $f_{{\bf k}}^{\dag}$ produces 
empty or vacant states,
Eq. (\ref{a_k}) shows that the binon is a kind of quantum
fluctuations of the tightly bound electron pairs.    
This observation is quite consistent with the crude belief
that the formation of the tightly bound electron pairs is a quantum 
fluctuation in the context of a continuum model and the
attractive Hubbard model\cite{ran1}. Here, we have mathematically
presented a possible and explicit description for such fluctuations.
By comparing the corresponding Fourier transforms we could obtain
the relationship between binons, vacanons and electrons:
$\sum_{{\bf q}}a_{{\bf q}+{\bf k}}^{\dag}f_{{\bf q}}=
\sum_{{\bf q}}c_{{\bf q}+{\bf k}\uparrow}^{\dag}c_{-{\bf q}
\downarrow}^{\dag}$ and $\sum_{{\bf q}}n_{{\bf q}-{\bf k}}^f =
\sum_{{\bf q}}[1- (n_{{\bf q}\uparrow} + n_{-{\bf q}+{\bf k}
\downarrow}) 
+ n_{{\bf q}}^a]$ where $n_{{\bf k}}^{z} = z_{{\bf k}}^{\dag}
z_{{\bf k}}$ ($z =a,f$). Equation (\ref{nfna=0}) gives
the condition $(1/M) \sum_{{\bf k}}{\hat {\cal F}}_{{\bf k}}=0$
with ${\hat {\cal F}}_{{\bf k}} = \sum_{{\bf q},{\bf q}'}
a_{{\bf q}+{\bf k}}^{\dag}a_{{\bf q}'+{\bf k}}f_{{\bf q}'}^{\dag}
f_{{\bf q}}$. The vacanon is supposed to obey $(1/M) \sum_{{\bf q}}
f_{{\bf k}+{\bf q}}^{\dag} f_{{\bf q}}|0\rangle = \delta_{{\bf k},0}
|0\rangle$. The total numbers of binons and vacanons satisfy
$N_a \in [0, N_{e}/2]$ and $N_f \in [0, M-N_{e}/2]$. To this end,
we could say that the bosons involved in the tightly bound electron
pairs could be identified as binons and vacanons which are 
actually responsible for the BEC phenomenon in the system. 
Although the concept of binons and vacanons is developed on the basis 
of the noninteracting tightly bound electron pairs, we
anticipate that it would be applicable to interacting electrons, 
bipolarons, and other systems.

\section*{The Attractive Hubbard Model with Strong Coupling}

It is known that the attractive Hubbard model in the strong 
coupling limit bears the form\cite{emery}  
\begin{eqnarray}
H =  \sum_{i,j}J_{ij} b_{i}^{\dag}b_{j} + \sum_{i,j}J_{ij}
b_{i}^{0} b_{j}^{0} - {\bar \mu} \sum_{i}n_{i} + const.
\end{eqnarray}
which is valid for arbitrary band filling, where ${\bar \mu}
= \mu + U/2$ with $\mu$ the chemical potential of electrons,
can be viewed as the effective chemical potential, and $J_{ij} = 
-2t_{ij}^2/|U|$. Hereafter we suppose $J_{ii}=0$. 
In terms of the terminology of binons and vacanons, 
this Hamiltonian can be rewritten as
\begin{eqnarray}
H =  \sum_{i,j}J_{ij} a_{i}^{\dag}a_{j}f_{j}^{\dag}f_{i}
+ \frac{1}{4} \sum_{i,j}J_{ij} (n_{i}^a n_{j}^a + n_{i}^f n_{j}^f
- 2 n_{i}^a n_{j}^f )  
+ {\bar \mu} \sum_{i} (n_{i}^f - n_{i}^a)
+ const.
\end{eqnarray}
Consequently,
the attractive Hubbard model with strong coupling is mapped onto a coupled
Bose system with binons and vacanons subject to the constraints
mentioned before. The first term denotes the hopping process of 
binons and vacanons, as shown in Fig.1, while the second sum shows the
interactions between binons and vacanons. In momentum space, the
Hamiltonian reads
\begin{eqnarray}
H = \frac{1}{M}
\sum_{{\bf k},{\bf q},{\bf q}'} \epsilon_{{\bf k}}
a_{{\bf q}+{\bf k}}^{\dag}a_{{\bf q}'+{\bf k}}
f_{{\bf q}'}^{\dag}f_{{\bf q}}  
-\frac{1}{4}\sum_{{\bf k}} \epsilon_{{\bf k}} (\rho_{{\bf k}}^{a}
-\rho_{{\bf k}}^{f})^2 +  {\bar \mu} \sum_{{\bf k}} 
(n_{{\bf k}}^{f}-n_{{\bf k}}^{a}) + const. 
\label{hamil-k}
\end{eqnarray}
where $\epsilon_{{\bf k}} = 
-(2t^2/|U|)\sum_{{\bf \delta}}\exp(i{\bf k}
\cdot {\bf \delta})$ (${\bf \delta}$ is the vector connecting
the nearest-neighbor lattice sites if we assume $t_{ij}=t$ for $i,j$
being nearest neighbors and zero otherwise), and $\rho_{{\bf k}}^{z} 
= (1/M) \sum_{{\bf q}} z_{{\bf q}}^{\dag}
z_{{\bf q}+{\bf k}}$ ($z =a,f$). One may observe that the BEC
phenonmenon in the attractive Hubbard model with strong coupling
can thus be explored on the basis of the above Hamiltonian in the restricted
Hilbert subspace where the binons and vacanons cannot occupy the
same site. Since this is a kind of interacting hard-core Bose gas, 
some approximations will be necessary. The details will be presented 
elsewhere.

We now consider the homogeneous continuum case at half-filling. 
The Bose field operators can be introduced through the 
following definitions:
\begin{eqnarray}
a_{i}^{\dag} = \int d{\bf r} \delta({\bf r}-{\bf r}_{i}) \psi_{a}^{\dag}
({\bf r}), ~~~\psi_{a}^{\dag}({\bf r}) = \sum_{i}
\delta({\bf r}-{\bf r}_{i})a_{i}^{\dag}. \nonumber \\
f_{i}^{\dag} = \int d{\bf r} \delta({\bf r}-{\bf r}_{i}) \psi_{f}^{\dag}
({\bf r}), ~~~\psi_{f}^{\dag}({\bf r}) = \sum_{i}
\delta({\bf r}-{\bf r}_{i})f_{i}^{\dag}, 
\end{eqnarray}
where $[\psi_{a,f}({\bf r}), \psi_{a,f}^{\dag}({\bf r}')] = \delta({\bf r}
-{\bf r}')$, $[\psi_{a,f}({\bf r}), \psi_{a,f}({\bf r}')] = 
[\psi_{a}({\bf r}), \psi_{f}({\bf r}')]=0$ and $\{\psi_{a,f}^{\dag}({\bf r})
\psi_{a,f}({\bf r})\}^2 = \psi_{a,f}^{\dag}({\bf r})
\psi_{a,f}({\bf r})$. It follows from Eq. (\ref{nfna=0}) that
$\psi_{a}^{\dag}({\bf r}) \psi_{a}({\bf r})\psi_{f}^{\dag}({\bf r})
\psi_{f}({\bf r})=0$.
The Hamiltonian becomes
\begin{eqnarray}
H &=& \int d{\bf r}d{\bf r}' \psi_{a}^{\dag}({\bf r}) \psi_{a}({\bf r}')
J({\bf r}-{\bf r}') \psi_{f}^{\dag}({\bf r}') \psi_{f}({\bf r}) 
\nonumber \\
&+& \frac14 \int d{\bf r}d{\bf r}' [\psi_{a}^{\dag}({\bf r}) 
\psi_{a}({\bf r}) J({\bf r}-{\bf r}') \psi_{a}^{\dag}({\bf r}') 
\psi_{a}({\bf r}') + \psi_{f}^{\dag}({\bf r}) 
\psi_{f}({\bf r}) J({\bf r}-{\bf r}') \psi_{f}^{\dag}({\bf r}') 
\psi_{f}({\bf r}') \nonumber \\
&-& 2 \psi_{a}^{\dag}({\bf r}) 
\psi_{a}({\bf r}) J({\bf r}-{\bf r}') \psi_{f}^{\dag}({\bf r}') 
\psi_{f}({\bf r}')] 
+ {\bar \mu} \int d{\bf r} [\psi_{f}^{\dag}({\bf r}) \psi_{f}({\bf r})
- \psi_{a}^{\dag}({\bf r}) \psi_{a}({\bf r})] + const.,
\label{hamil}
\end{eqnarray} 
where $J({\bf r}-{\bf r}') = \sum_{i,j} \delta({\bf r}-{\bf r}_{i})
\delta({\bf r}'-{\bf r}_{j}) J_{ij}$. 
From the equations of motion for the 
field operators $\psi_{a}({\bf r}, t)$ and $\psi_{f}({\bf r}, t)$
we obtain
\begin{eqnarray}
i\hbar \frac{\partial \Psi ({\bf r}, t)}{\partial t} = h_{eff} ({\bf r}, t)
\Psi ({\bf r}, t)
\label{motion}
\end{eqnarray}
with
\begin{eqnarray}
& & \Psi ({\bf r}, t) = \biggl(\matrix{\psi_{a}({\bf r},t) \cr 
\psi_{f}({\bf r},t)}\biggr), \\
& & h_{eff} ({\bf r}, t) = \biggl(\matrix{\zeta({\bf r}, t) -{\bar \mu} &
\xi ({\bf r}, t) \cr \xi^{*}({\bf r}, t) & -[\zeta({\bf r}, t) -{\bar 
\mu}] \cr}\biggr), \label{heff} \\
& & \xi ({\bf r}, t) = \int d{\bf r}' \psi_{f}^{\dag}({\bf r}',t) 
J({\bf r}-{\bf r}') \psi_{a}({\bf r}',t), \\
& & \zeta ({\bf r}, t) = \frac12 
\int d{\bf r}' [\psi_{a}^{\dag}({\bf r}',t) 
J({\bf r}-{\bf r}') \psi_{a}({\bf r}',t) - 
\psi_{f}^{\dag}({\bf r}',t) 
J({\bf r}-{\bf r}') \psi_{f}({\bf r}',t)],
\end{eqnarray}
where we have noticed that $J({\bf r}-{\bf r}')$ is a real function.
One may observe that Eq. (\ref{motion}) is nothing but the Schr\"odinger
equation with the effective, time-dependent local Hamiltonian 
$h_{eff}({\bf r}, t)$, which implies that $[\Psi ({\bf r}), H] =
h_{eff} ({\bf r})\Psi ({\bf r})$.
The formal solution of $\Psi ({\bf r}, t)$ can be written as
\begin{eqnarray}
\Psi ({\bf r}, t) = \Psi ({\bf r}, t_{0}) e^{-\frac{i}{\hbar}
\int_{t_{0}}^{t} h_{eff}({\bf r}, t')dt'}.
\end{eqnarray}
It turns out that the Hamiltonian (\ref{hamil})
can be equivalently rewritten as
\begin{eqnarray}
H = \frac12 \int d{\bf r} \Psi^{\dag} ({\bf r}) (h_{eff}({\bf r})
-{\bar \mu}) \Psi ({\bf r})
+ const.,
\label{hamil-e}
\end{eqnarray}
where use has been made of $J(0)=0$ and $J({\bf r}-{\bf r}') =
J({\bf r}'-{\bf r})$. It should be noted that $\Psi ({\bf r})$ itself
does not fulfil the bosonic commutation relations but its elements do. 
If one makes proper approximations on $h_{eff}({\bf r})$, one could 
get the Gross-Pitaevskii (GP)-like equation\cite{gp} which is known 
to be capable of giving a better description 
on the condensate in dilute Bose gas near zero temperature. 
In this sense, Eq. (\ref{motion}) can be
regarded as a generalized GP equation for the system under interest.
Like the Bogoliubov-de Gennes equation in superconductors, we may here
linearize the Hamiltonian by replacing $\xi ({\bf r})$ and 
$\zeta ({\bf r})$ by their averages ${\bar \xi({\bf r})} \equiv
\langle \xi ({\bf r}) \rangle$ and ${\bar \zeta({\bf r})} \equiv
\langle \zeta ({\bf r}) \rangle$, respectively. Then, we can
seek for the stationary solution of interest of the form
$h_{eff} ({\bf r}) \Psi ({\bf r}) \equiv E({\bf r}) 
\Psi ({\bf r})$, with
\begin{eqnarray}
E({\bf r}) = \sqrt{({\bar \zeta({\bf r})}-{\bar \mu})^2 
+ |{\bar \xi({\bf r})}|^2}.
\label{Er}
\end{eqnarray}
Within this linearizing approximation, 
the attractive Hubbard model in the strong coupling limit
is mapped onto the system of noninteracting binons 
and vacanons described by Eq. (\ref{hamil-e}), subject to the constraint
that the binons and vacanons cannot occupy the same site.
For the $N_a$-binon state of the form 
\begin{eqnarray}
|\phi ({\bf k})\rangle = \frac{1}{\sqrt{N_{a}!}} [a^{\dag}({\bf k})]^{N_a}
|0\rangle,
\label{phi}
\end{eqnarray}
where $a^{\dag}({\bf k})$ is the Fourier transform of 
$\psi_{a}^{\dag}({\bf r})$,  
one may verify that $|\phi ({\bf k})\rangle$ is an eigenstate
of the linearized Hamiltonian $H$.
Recall that $\langle \phi ({\bf  k}) | \phi ({\bf  k}') \rangle = 
\delta_{{\bf k}{\bf k}'}$, $\int d{\bf k}| \phi ({\bf  k}) \rangle
\langle \phi ({\bf  k}) | =1$.  On the other hand, we can prove that
\begin{eqnarray}
\langle \phi ({\bf k})| \psi_{a}^{\dag}({\bf r})\psi_{a}({\bf r}')
|\phi ({\bf k}) \rangle = \frac{N_{a}}{M} e^{-i {\bf k} \cdot (
{\bf r}-{\bf r}')}.
\end{eqnarray}
If $N_{a}$ is of the order $O(M)$, then $\lim_{|{\bf r}-{\bf r}'| \to
\infty} \langle \phi ({\bf k})| \psi_{a}^{\dag}({\bf r})\psi_{a}({\bf r}')
|\phi ({\bf k}) \rangle = O(1) \neq 0$, implying the existence of ODLRO in
the one-particle reduced density matrix, $\rho_{1}$, of the binons 
in the state $|\phi ({\bf k})\rangle$. While
the existence of ODLRO in $\rho_{1}$ in Bose systems implies the
macroscopic occupation of bosons at a certain quantum state (i.e., 
the BEC), as discussed clearly by Penrose and Onsager\cite{penrose},
it suggests that the existence of the BEC of binons in the attractive 
Hubbard model in the strong coupling limit be possible. Consequently, 
we could understand qualitatively the whole crossover behavior
from BCS to BEC in the attractive Hubbard model with varying the
coupling constant as follows. In the weak coupling limit, the system 
should be in the BCS regime characterizing the formation of Cooper
pairs; when the coupling is smoothly increasing to intermediate 
values, the system goes into the crossover regime in which Cooper 
pairs are getting broken and the charge fluctuations somehow become
dominant; when the coupling becomes very strong, the system enters 
into the BEC regime in which the binons and vacanons play an important 
role. This understanding is qualitatively consistent 
with the results obtained
in Ref.\cite{ran1}. To this end, we have shown that the bosonic
quasiparticles, binons and vacanons, play an important role in the
tightly bound electron systems, which may be in fact 
responsible for the BEC
of the system. Certainly, more analytical and numerical 
works are needed to confirm this observation.

\section*{Concluding remarks}

We have shown in this paper that the bosonic degree of freedom 
of the tightly bound on-site electron pairs could be separated 
as Schwinger bosons which are two Bose quasiparticles 
named as binon and vacanon. This can be implemented by projecting 
the whole Hilbert space into the
Hilbert subspace spanned by states of binons and vacanons subject 
to a constraint that they cannot occupy the same site. 
We argue that a binon, carrying the same charge and spin as the
tightly bound on-site electron pairs, is actually a kind of quantum 
fluctuations of electron pairs, and a vacanon, carrying both charge 
and spin zero, corresponds to a vacant state. 
These two bosonic quasiparticles may be responsible for the  
BEC of the system consisting of the tightly bound electron pairs. 
These concepts were also applied to the attractive Hubbard model with 
strong coupling in continuum case, showing that the binon-vacanon 
picture for describing the BEC phenomenon in this special system
is useful. We expect that our arguments could be applicable to
the correlated electron systems.

We close this paper by discussing briefly the relevance of our 
argument to the existing theories concerning the BEC of electron 
pairs.

(1) The crossover picture\cite{ue1} for high-temperature 
superconducting cuprates is based on such an observation 
that the superconducting transition temperature
is proportional to the superfluid density in the underdoped regime which
is supposed to be attributed to the BEC of 
pre-formed singlet pairs (or bosons). 
If this picture is sound, the dispersion relation of such bosons should
depend linearly on momentum in two dimensions 
(see Appendix). The pre-formed 
bosons can be interpreted as the binons introduced in the present
paper, and the superfluid density could be interpreted as $n_{a}$.
In terms of this hypothesis, the London penetration depth obeys
$\lambda^2(0)/\lambda^2(T) = 1-(T/T_c)$, which somewhat deviates 
quantitatively from the standard BCS result for a $d_{x^2-y^2}$-wave 
superconductor\cite{xiang}: $\lambda^2(0)/\lambda^2(T) \approx 1-0.65
(T/T_c)$.

(2) In the crossover theory from the BCS to the BEC regimes in electron 
systems, the associated bosons could not be the tightly bound electron 
pairs. Instead, they may be understood as a kind of quantum 
fluctuations, e.g.,
the binons and vacanons specified in this paper.

(3) In the boson-fermion theory of superconductivity\cite{jan}, the small
bipolarons were considered as hard core bosons on a lattice, and were
simply replaced by Bose operators in the formalism. Now, we have offered 
a reasonable explanation for such a replacement, namely, the hard core
bosons can be reinterpreted as the binons and vacanons
introduced in this paper\cite{chak}.
 
(4) In Schafroth's superconducting theory\cite{schaf} 
which is being considered as a 
mechanism of bipolaronic theory of superconductivity\cite{alex}, 
superconductivity
arises from the BEC of charged ideal Bose gas. Based on our result, 
those charged bosons cannot be directly the tightly bound electron pairs,  
while they can be interpreted as the Schwinger bosons (e.g. binons). 

\acknowledgments

One of the authors (GS) is grateful to the Department of Applied Physics,
Science University of Tokyo, for the warm hospitality, and to the Japan 
Society for the Promotion of Science (JSPS) for support. 
This work has also been supported by the CREST (Core Research for 
Evolutional Science and Technology) of the Japan
Science and Technology Corporation (JST).

\section*{appendix}

In this appendix, for reader's convenience we write down a general
formula of the BEC transition temperature for an isotropic system 
of noninteracting bosonic quasiparticles with
dispersion $\epsilon_{{\bf p}} =  \gamma |{\bf p}|^{\alpha}$ where
${\bf p}$ is the momentum, $\alpha$ and $\gamma$ are positive 
constants which
are chosen in terms of the dimension of energy by\cite{suzuki}
\begin{eqnarray}
T_{c} = \frac{(2^{1-\frac{1}{d}} \pi^{\frac12} 
\hbar)^{\alpha}
\gamma}{k_{B}}[\frac{\Gamma (\frac{d}{2}) n \alpha}{\Gamma 
(\frac{d}{\alpha})
\zeta (\frac{d}{\alpha})}]^{\frac{\alpha}{d}},
\nonumber
\end{eqnarray}
where $n = \langle {\hat N} \rangle/M$ is 
the density of the bosonic quasiparticles, $d$
is the dimensionality, $\Gamma (x)$ is the gamma function,
and $\zeta (x) = 
\sum_{n=0}^{\infty} n^{-x}$ $(x>1)$  is the Riemann zeta function.
 The condensed fraction of
bosons, $n_{0}$, can be expressed by $n_{0} = n [1 - (T/T_{c})^{
d/\alpha}]$. The Bernouli equation becomes\cite{su} $PV = 
\frac{\alpha}{d} E$ with $E$ the internal energy. The specific heat
for $T<T_c$ can be obtained by
\begin{eqnarray}
c_{v} = \frac{(\frac{d}{\alpha}+1)k_{B}}{2^{d-1}\pi^{d/2}\hbar^{d}
\alpha}\frac{\Gamma(\frac{d}{\alpha}+1)\zeta(\frac{d}{\alpha}+1)}
{\Gamma(\frac{d}{2})} \biggl(\frac{k_{B}T}{\gamma}\biggr)^{\frac{d}
{\alpha}}.
\nonumber
\end{eqnarray}
For $T>T_{c}$, one must consider the effect of the temperature-dependence
of the chemical potential, $\mu = \mu(T)$, but keep the density fixed. At
high temperatures, $c_{v} \to \frac{d}{\alpha}k_{B}n$.

\begin{figure}
\narrowtext
\epsfxsize=\linewidth
%\vspace{2mm}
\center{\mbox{\hspace{2mm}\psfig{figure=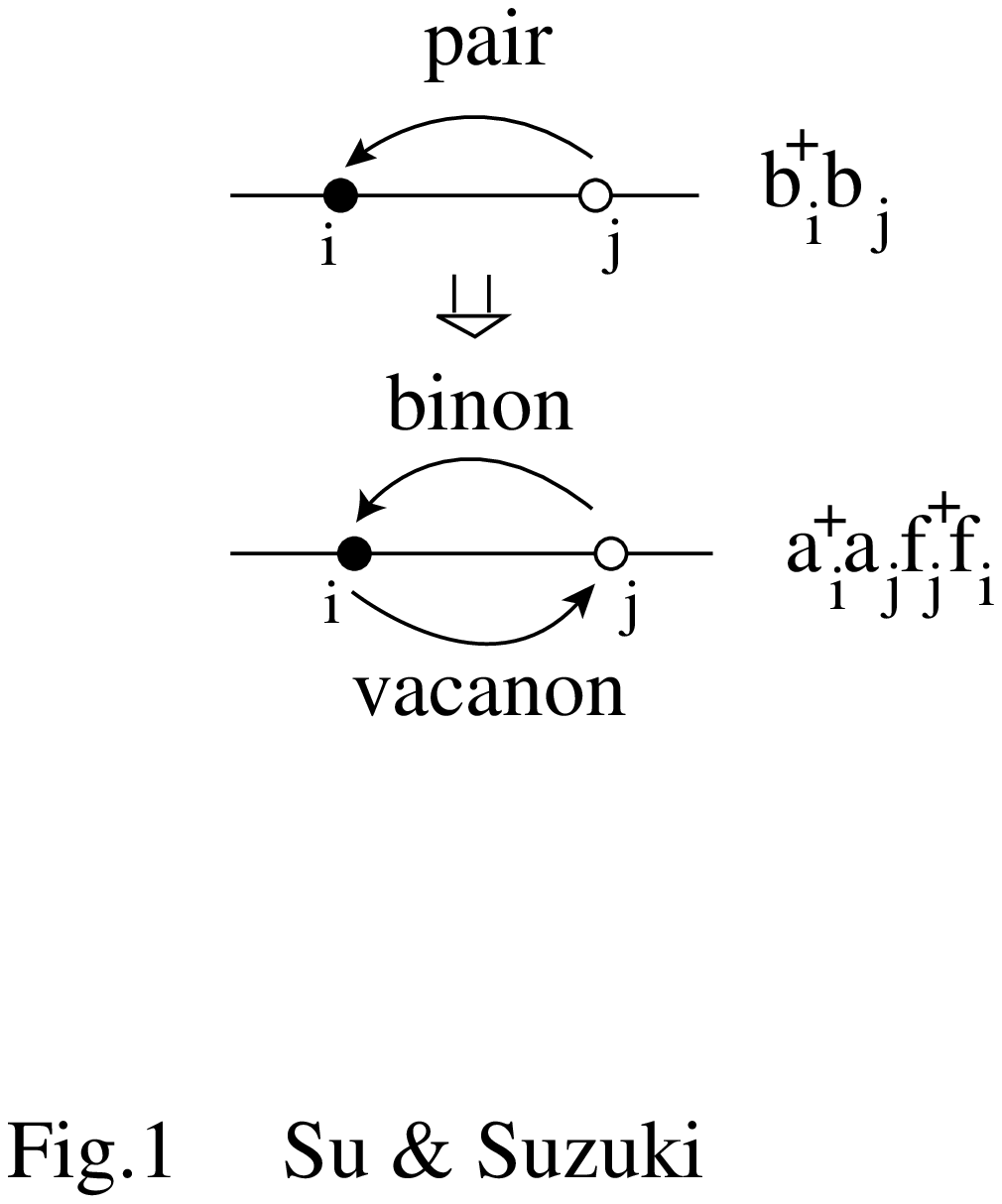,width=30mm,height=35mm}}}
\caption{The hopping process of a tightly bound 
electron pair is associated
with two excitations (i.e. binon and vacanon) with opposite hopping
processes. }
%\label{fig1}
\end{figure}

\begin{references}
\bibitem{note00}In some literature the tightly bound local pairs were
considered as a weakly nonideal Bose gas under the assumption of
the very dilute limit. We believe that the approximation is not so
good even for this limit, and especially, the approximations 
make the physical meaning of the bosons invloved in the system ambiguous.
\bibitem{note0}Without loss of the generality, 
we here confine ourselves to consider
only the on-site electron pairs. The intersite pairs, 
owing to the complexity, are left for studying elsewhere.
\end{references}
\end{document}